\documentclass[10pt,conference]{IEEEtran}
\usepackage{psfrag}
\usepackage{subfigure}
\usepackage{cite}

\usepackage{array,multirow}
\usepackage{graphicx}  
\usepackage{psfrag}    
\usepackage{subfigure} 
\usepackage{amsmath}   
\usepackage{amssymb}   
\usepackage{eucal}     

\usepackage{pxfonts}
\usepackage{dsfont}

\newtheorem{theorem}{\textbf{Theorem}}
\newtheorem{corollary}{\textbf{Corollary}}

\newtheorem{definition}{\textbf{Definition}}

\newtheorem{remark}{Remark}
\newcommand{\dv}{\mathbf} 
\newcommand{\mc}{\mathcal} 
\newcommand{\qed}{\hfill \ensuremath{\Box}}

\DeclareMathAlphabet{\eurm}{U}{eur}{m}{n}
\DeclareMathAlphabet{\mathbsf}{OT1}{cmss}{bx}{n}
\DeclareMathAlphabet{\mathssf}{OT1}{cmss}{m}{sl}
\DeclareMathAlphabet{\mathcsf}{OT1}{cmss}{sbc}{n}

\DeclareSymbolFont{bsfletters}{OT1}{cmss}{bx}{n}  
\DeclareSymbolFont{ssfletters}{OT1}{cmss}{m}{n}
\DeclareMathSymbol{\bsfGamma}{0}{bsfletters}{'000}
\DeclareMathSymbol{\ssfGamma}{0}{ssfletters}{'000}
\DeclareMathSymbol{\bsfDelta}{0}{bsfletters}{'001}
\DeclareMathSymbol{\ssfDelta}{0}{ssfletters}{'001}
\DeclareMathSymbol{\bsfTheta}{0}{bsfletters}{'002}
\DeclareMathSymbol{\ssfTheta}{0}{ssfletters}{'002}
\DeclareMathSymbol{\bsfLambda}{0}{bsfletters}{'003}
\DeclareMathSymbol{\ssfLambda}{0}{ssfletters}{'003}
\DeclareMathSymbol{\bsfXi}{0}{bsfletters}{'004}
\DeclareMathSymbol{\ssfXi}{0}{ssfletters}{'004}
\DeclareMathSymbol{\bsfPi}{0}{bsfletters}{'005}
\DeclareMathSymbol{\ssfPi}{0}{ssfletters}{'005}
\DeclareMathSymbol{\bsfSigma}{0}{bsfletters}{'006}
\DeclareMathSymbol{\ssfSigma}{0}{ssfletters}{'006}
\DeclareMathSymbol{\bsfUpsilon}{0}{bsfletters}{'007}
\DeclareMathSymbol{\ssfUpsilon}{0}{ssfletters}{'007}
\DeclareMathSymbol{\bsfPhi}{0}{bsfletters}{'010}
\DeclareMathSymbol{\ssfPhi}{0}{ssfletters}{'010}
\DeclareMathSymbol{\bsfPsi}{0}{bsfletters}{'011}
\DeclareMathSymbol{\ssfPsi}{0}{ssfletters}{'011}
\DeclareMathSymbol{\bsfOmega}{0}{bsfletters}{'012}
\DeclareMathSymbol{\ssfOmega}{0}{ssfletters}{'012}

\newcommand{\calS}{{\mathcal{S}}}

\newcommand{\calX}{{\mathcal{X}}}
\newcommand{\calY}{{\mathcal{Y}}}

\begin{document}
{\fontencoding{OT1}\fontsize{9.4}{11.25pt}\selectfont
\title{Bounds on the Capacity of the Relay Channel with Noncausal State Information at Source\\}

\author{Abdellatif Zaidi$\:^{\dagger}$ \qquad Shlomo Shamai (Shitz)$\:^{\ddagger}$ \qquad Pablo Piantanida$\:^{\nmid}$ \qquad Luc Vandendorpe$\:^{\dagger}$\vspace{0.3cm}\\
$^{\dagger}$ \'{E}cole Polytechnique de Louvain, Universit\'{e} catholique de Louvain, LLN-1348, Belgium\\
$^{\ddagger}$ Department of EE, Technion-Israel Institute of Technology, Haifa, Israel\\
$^{\nmid}$ Department of Telecommunications, SUPELEC, 91192 Gif-sur-Yvette, France\\
\{abdellatif.zaidi,luc.vandendorpe\}@uclouvain.be\\
sshlomo@ee.technion.ac.il, pablo.piantanida@supelec.fr
}

\maketitle

\begin{abstract}
We consider a three-terminal state-dependent relay channel with the channel state available non-causally at only the source. Such a model may be of interest for node cooperation in the framework of cognition, i.e., collaborative signal transmission involving cognitive and non-cognitive radios. We study the capacity of this communication model. One principal problem in this setup is caused by the relay's not knowing the channel state. In the discrete memoryless (DM) case, we establish lower bounds on channel capacity. 
For the Gaussian case, we derive lower and upper bounds on the channel capacity. The upper bound is strictly better than the cut-set upper bound. We show that one of the developed lower bounds comes close to the upper bound, asymptotically, for certain ranges of rates. 
\end{abstract}

\section{Introduction}\label{secI}

We consider a three-terminal state-dependent relay channel (RC) in which, as shown in Figure~\ref{StateDependentDiscreteMemorylessRelayChannel}, the source wants to communicate a message $W$ to the destination through the state-dependent RC in $n$ uses of the channel, with the help of the relay. The channel outputs $Y_2$ and $Y_3$ for  the relay and the destination, respectively, are controlled by the channel input $X_1$, the relay input $X_2$ and the channel state $S$, through a given memoryless probability law $W_{Y_2,Y_3|X_1,X_2,S}$. The channel state $S$ is generated according to  a given memoryless probability law $Q_S$. It is assumed that the channel state is known, noncausally, to only the source. The destination estimates the message sent by the source from the received channel output. In this paper we study the capacity of this communication system. We refer to this model as \textit{state-dependent RC with informed source}.

\begin{figure}[htpb]
\centering
        \includegraphics[width=\linewidth]{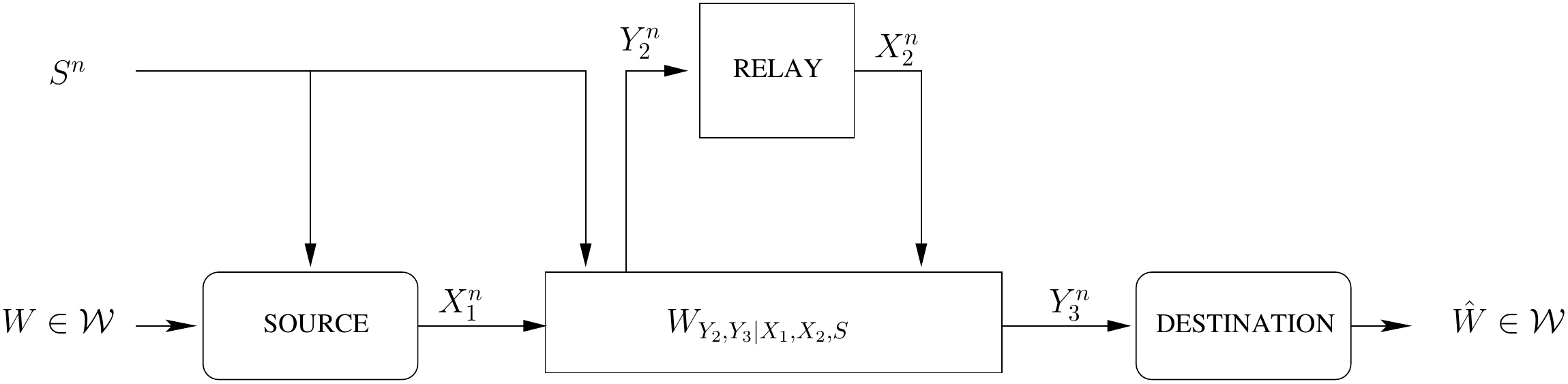}
\caption{Relay channel with state information $S^n$ available non-causally at only the source.}
\label{StateDependentDiscreteMemorylessRelayChannel}
\end{figure}

The state-dependent multiaccess channel (MAC) with only one informed encoder and degraded message sets is considered in \cite{KL07a,SBSV07a}; and the state-dependent relay channel (RC) with only informed relay is considered in \cite{ZKLV08b}. For all these models, the authors develop non-trivial outer or upper bounds that permit to characterize the rate loss due to not knowing the state at the uninformed encoders. Key feature to the development of these outer or upper bounds is that, in all these models, the uninformed encoder not only does not know the channel state but, also, can learn no information about it. 

The model for the RC with informed source that we study in this paper seemingly exhibits some similarities with the RC with informed relay considered in \cite{ZKLV08b}, and it also connects with the MAC with asymmetric CSI and degraded message sets considered in \cite{KL07a,SBSV07a,ZKLV09a} and with the MAC with two states considered in \cite{PKEZ07}. However, establishing a non-trivial upper bound for the present model is more involved, comparatively. Partly, this is because, here, the uninformed encoder is also a receiver; and, so, it can potentially get some information about the channel states from directly observing its output. That is, at time $i$, the input $X_{2,i}$ of the relay can potentially depend on the channel states through $Y_2^{i-1}=(Y_{2,1},\hdots,Y_{2,i-1})$. Further, since, for $j=1,\hdots,i-1$, $Y_2^j$ may depend on the channel states in a non-causal manner (through the source codeword $X_{1,j}(W,S^n)$), and not only through the current state $S_i$, so does the input of the relay, potentially.

Establishing good lower bounds for the present model is also a non-easy task, due to the asymmetry caused by knowing the state at only the source. In this paper, we establish two lower bounds on the capacity of the state-dependent RC with informed source, for both discrete memoryless (DM) and memoryless Gaussian cases. For the Gaussian case, we also establish a non-trivial upper bound that is strictly better than the max-flow min cut or cut-set upper bound. Our lower bounds exploit ideas that are inherently different; and, so, their comparison helps providing right guidance towards the appropriate design. We obtain the first lower bound by a coding scheme in which the source describes the known state to the relay and destination \textit{ahead of time}, in addition to sending the information message. The relay performs collaborative binning against the learned state, through a combined binning and decode-and-forward (DF) scheme. We obtain the second lower bound by a coding scheme in which, rather than the channel state itself, the source describes to the relay the appropriate input that the relay would send had the relay known the channel state. The relay then simply guesses this input and sends it in the appropriate block. The lower bound obtained with this scheme achieves close to optimal for some special cases. 

We note that the lower bounds that we develop in this paper are tailored to primarily overcome the state asymmetry; and, so, they perform well in the situations in which classic DF for RC without state is suitable. Other interesting achievability results which perform well in other situations can be found in \cite{AMA09,KS09,ZV09b}.

\section{System Model and Definitions}\label{secII}
As shown in Figure~\ref{StateDependentDiscreteMemorylessRelayChannel}, we consider a state-dependent relay channel  denoted by $W_{Y_2,Y_3|X_1,X_2,S}$ whose outputs $Y_2 \in \mc Y_2$ and $Y_3 \in \mc Y_3$ for the relay and the destination, respectively, are controlled by the channel inputs $X_1 \in \mc X_1$ from the source and $X_2 \in \mc X_2$ from the relay, along with a random state parameter $S \in \mc S$. It is assumed that the channel state $S_i$ at time instant $i$ is independently drawn from a given distribution $Q_S$ and the channel states $S^n$ are non-causally known at the source.

The source wants to transmit a message $W$ to the destination with the help of the relay, in $n$ channel uses. The message $W$ is assumed to be uniformly distributed over the set $\mc W=\{1,\hdots,M\}$. The information rate $R$ is defined as $n^{-1}\log M$ bits per transmission.

\noindent An $(M,n)$ code for the state-dependent relay channel with informed source consists of an encoding function at the source
$$\phi_1^n:~~\{1,\hdots,M\}\times \calS^n \rightarrow \mc X_1^n,$$ a sequence of encoding functions at the relay
$$\phi_{2,i}:~~ \calY_{2,1}^{i-1} \rightarrow \calX_2,$$
for $i=1,2,\ldots,n,$ and a decoding function at the destination
$$\psi^n: ~~ \mc Y_3^n \rightarrow \{1,\hdots,M\}.$$

Let a $(M,n)$ code be given. The sequences $X_{1}^n$ and $X_{2}^n$ from the source and the relay, respectively, are transmitted across a state-dependent relay channel modeled as a memoryless conditional probability distribution $W_{Y_2,Y_3|X_1,X_2,S}$. The joint probability mass function on ${\mc W}{\times}{\mc S^n}{\times}{\mc X^n_1}{\times}{\mc X^n_2}{\times}{\mc Y^n_2}{\times}{\mc Y^n_3}$ is given by
\begin{align}
P(w,s^n,x^n_1,x^n_2,y^n_2,y^n_3) = &P(w)\prod_{i=1}^{n}Q_S(s_i)P(x_{1,i}|w,s^n)P(x_{2,i}|y^{i-1}_2)\nonumber\\
&{\cdot}W_{Y_2,Y_3|X_1,X_2,S}(y_{2,i},y_{3,i}|x_{1,i},x_{2,i},s_i).
\end{align}

The destination estimates the message sent by the source from the channel output $Y_{3}^n$. The average probability of error is defined as $P_e^n = \mathbb{E}_{S}\big[\mathrm{Pr}\big(\psi^n(Y_{3}^n)\neq W|S^n=s^n\big)\big].$

An  $(\epsilon,n,R)$ code for the state-dependent RC with informed source is an $(2^{nR},n)-$code $(\phi_1^n,\phi_2^n,\psi^n)$ having average probability of error $P_e^n$ not exceeding $\epsilon$.

A rate $R$ is said to be achievable if there exists a sequence of $(\epsilon_n,n,R)-$codes with $\lim_{n \rightarrow \infty} \epsilon_n=0$. The capacity $\mc C$ of the state-dependent RC with informed source is defined as the supremum of the set of achievable rates.

Due to space limitation, the results of this paper are either outlined only or mentioned without proofs. Detailed proofs can be found in \cite{ZS10}.

\section{The Discrete Memoryless RC with Informed Source}\label{secIII}

In this section, we assume that the alphabets $\calS$, $\calX_1$, $\calX_2$, $\calY_2$, $\calY_3$ in the model are all discrete and finite.



\vspace{0.1cm}

\begin{theorem}\label{Theorem__LowerBound1__DiscreteChannel}
The capacity of the discrete memoryless state-dependent relay channel with informed source is lower bounded by
\begin{align}
R^{\text{lo}} = \max \: \min\: \{&I(U;Y_2|V,\hat{S}_R)-I(U;S,\hat{S}_D|V,\hat{S}_R),\nonumber\\
&I(U,V;Y_3|\hat{S}_D)-I(U,V;S,\hat{S}_R|\hat{S}_D)\}
\label{AchievableRate__Theorem__LowerBound1__DiscreteChannel}
\end{align}
subject to the constraints
\begin{subequations}
\begin{align}
I(S;\hat{S}_R) &\leq I(U_R;Y_2,\hat{S}_R|U,V)-I(U_R;S,\hat{S}_R,\hat{S}_D|U,V)\\
I(S;\hat{S}_D) &\leq I(U_D;Y_3,\hat{S}_D|U,V)-I(U_D;S,\hat{S}_R,\hat{S}_D|U,V)\nonumber\\
&+[I(U;Y_3,\hat{S}_D|V)-I(U;S,\hat{S}_R,\hat{S}_D|V)]_{-}\\
I(S;\hat{S}_R,\hat{S}_D) + I(\hat{S}_R;\hat{S}_D) &\leq I(U_R;Y_2,\hat{S}_R|U,V)-I(U_R;S,\hat{S}_R,\hat{S}_D|U,V)\nonumber\\
&+I(U_D;Y_3,\hat{S}_D|U,V)-I(U_D;S,\hat{S}_R,\hat{S}_D|U,V)\nonumber\\
&+[I(U;Y_3,\hat{S}_D|V)-I(U;S,\hat{S}_R,\hat{S}_D|V)]_{-}\nonumber\\
&-I(U_R;U_D|U,V,S,\hat{S}_R,\hat{S}_D)
\label{Constraints__AchievableRate__Theorem__LowerBound1__DiscreteChannel}
\end{align}
\end{subequations}
where $[x]_{-} \triangleq \min(x,0)$, and the maximization is over all joint measures on ${\mc S}\times\hat{\mc S}_R\times\hat{\mc S}_D\times{\mc U}_R\times{\mc U}_D\times{\mc U}\times{\mc V}\times{\mc X}_1\times{\mc X}_2\times{\mc Y}_2\times{\mc Y}_3$ of the form
\begin{align}
&P_{S,\hat{S}_R,\hat{S}_D,U_R,U_D,U,V,X_1,X_2,Y_2,Y_3} = Q_SP_{\hat{S}_R,\hat{S}_D|S}P_{V|\hat{S}_R}P_{U|V,S,\hat{S}_R,\hat{S}_D}P_{U_R,U_D|V,U,S,\hat{S}_R,\hat{S}_D}\nonumber\\
&\qquad P_{X_1|U_R,U_D,U,V,S,\hat{S}_R,\hat{S}_D}P_{X_2|V,\hat{S}_R}W_{Y_2,Y_3|X_1,X_2,S}.
\label{Measure__AchievableRate__Theorem__LowerBound1__DiscreteChannel}
\end{align}
and satisfying
\begin{align}
I(V;Y_3,\hat{S}_D) - I(V;\hat{S}_R) > 0.
\end{align}
\end{theorem}
\vspace{0.3cm}

\begin{remark}\label{remark1}
The intuition for the coding scheme which we use to establish the lower bound in Theorem~\ref{Theorem__LowerBound1__DiscreteChannel} is as follows. Had the relay known the state, the source and the relay could implement collaborative binning against that state for transmission to the destination \cite{KSS04}. Since the source knows the state of the channel non-causally, it can transmit a description of it to the relay \textit{ahead of time}. The relay recovers the state (with a certain distortion), and  then utilizes it in the relevant block through a collaborative binning scheme. The hope is that the benefit that the source can get from being assisted by a more capable relay largely compensates the loss caused by the source's spending some of its resources to make the relay learn the state.

\noindent In general, it may also turn to be useful to send a dedicated description of the state to the destination. The destination utilizes the recovered state as side information at the receiver. The coding scheme that we employ to establish the lower bound in Theorem~\ref{Theorem__LowerBound1__DiscreteChannel} uses block Markov encoding. In each block $i$, in addition to its message, the source also sends a two-layer description of the state $\dv s[i+2]$ to the relay and destination; a description $\hat{\dv s}_R[\iota_{Ri}]$ of $\dv s[i+2]$ intended to be recovered at the relay and a description $\hat{\dv s}_D[\iota_{Di}]$ of $\dv s[i+2]$ intended to be recovered at the destination. The two layers are possibly correlated. (In Remark~\ref{remark2}, we will comment on the delay of two blocks needed here). The relay guesses the source's message $w_i$ and the individual state description $\hat{\dv s}_R[\iota_{Ri}]$ dedicated to it from the source transmission and the previously recovered state description $\hat{\dv s}_R[\iota_{Ri-2}]$. It will then utilize the new state estimate as non-causal state at the encoder for collaborative source-relay binning in block $i+2$, through a combined decode-and-forward and Gelf'and-Pinsker binning. The destination guesses the source's message $w_{i-1}$ sent cooperatively by the source and relay and the individual state description $\hat{\dv s}_D[\iota_{Di-1}]$ which is dedicated to it from its output $(\dv y_3[i-1],\dv y_3[i])$ and the previously recovered state $(\hat{\dv s}_D[\iota_{Di-3}],\hat{\dv s}_D[\iota_{Di-2}])$, using a window-decoding method.
\end{remark}

\begin{remark}\label{remark2}
The source sends the descriptions intended to the relay and destination \textit{two blocks} ahead of time. That is, at the beginning of block $i$ the source describes the state vector $\dv s[i+2]$ to the relay and destination. While one block delay is sufficient to describe the state to the relay, a minimum of two blocks is necessary for the state reconstruction at the destination because of the used window decoding technique. 
\end{remark}

We can generalize Theorem~\ref{Theorem__LowerBound1__DiscreteChannel} by allowing the source to also send a common description of the state which is intended to be recovered at both the relay and the destination (see \cite{ZS10}). 

The following theorem provides a lower bound on the capacity of the state-dependent general discrete memoryless RC with informed source.

\vspace{0.1cm}

\begin{theorem}\label{Theorem__LowerBound2__DiscreteChannel}
The capacity of the discrete memoryless state-dependent relay channel with informed source is lower bounded by
\begin{align}
R^{\text{lo}} = \max \: \min\: \{&I(U,U_R;Y_3)-I(U,U_R;S),\nonumber\\
&I(U,U_R;Y_2,\hat{X})-I(U,U_R;S)-I(X;\hat{X})\}
\label{AchievableRate__Theorem__LowerBound2__DiscreteChannel}
\end{align}
subject to the constraint
\begin{align}
I(X;\hat{X}) &< I(U_R;Y_2,\hat{X}|U)-I(U_R;S|U)+[ I(U;Y_2,\hat{X})-I(U;S)]_{-}
\label{Constraint__AchievableRate__Theorem__LowerBound2__DiscreteChannel}
\end{align}
where $[x]_{-}=\min(x,0)$, and the maximization is over all joint measures on ${\mc S}\times{\mc U}\times{\mc U_R}\times{\mc X}_1\times{\mc X}_2\times{\mc X}\times\hat{\mc X}\times{\mc Y}_2\times{\mc Y}_3$ of the form
\begin{align}
&P_{S,U,U_R,X_1,X_2,X,\hat{X},Y_2,Y_3}\nonumber\\
&\quad = Q_SP_{U|S}P_{U_R|U,S}P_{X_1|U_R,U,S}P_{X|U,S}P_{\hat{X}|X}\mathds{1}_{X_2=\hat{X}}W_{Y_2,Y_3|X_1,X_2,S}.
\label{Measure__AchievableRate__Theorem__LowerBound2__DiscreteChannel}
\end{align}
\end{theorem}
\vspace{0.3cm}

\begin{remark}\label{remark5}
The rationale for the coding scheme which we use to obtain the lower bound in Theorem~\ref{Theorem__LowerBound2__DiscreteChannel} is as follows. With DF relaying, had the relay known the state then in each block the relay generates its input using the source transmission in the previous block and the state that controls the channel in the current block, as in \cite{KSS04}. For our model, the source knows what cooperative information, i.e., part of the message, the relay would send in each block. It also knows the state sequence that corrupts the transmission in that block. It can then generate the appropriate relay input vector that the relay would send had the relay known the state. The source can send this vector to the relay \textit{ahead of time}, and if the relay can estimate it to high accuracy, then collaborative source-relay binning in the sense of \cite{KSS04} is readily realized for transmission from the source and relay to the destination.
\end{remark}

\textbf{Outline of Proof:} A block Markov encoding with $B+1$ blocks is used.  Let us denote by $\dv x[k]$ the relay input carrying message $w_k \in [1,2^{nR}]$ that the relay would send in block $k$ had the relay known the state $\dv s[k]$, assuming DF relaying, with $k=2,\hdots,B+1$. Let us now consider transmission in two adjacent blocks $i$ and $i+1$. In the beginning of block $i$, the source sends information message $w_i$ of the current block, and, in addition, describes to the relay the input $\dv x[i+1]$ that the relay would send in the next block $i+1$ had the relay known the state $\dv s[i+1]$. Let $\hat{\dv x}[m_i]$ be a description of $\dv x[i+1]$. The source generates its input $\dv x_1[i]$ using two auxiliary codewords that are superimposed, a codeword $\dv u_R[i]$ that carries the index $m_i$ on top of a codeword $\dv u[i]$ that carries message $w_i$. Both codewords are selected using binning against the state $\dv s[i]$ that controls transmission in the current block $i$. The vector $\dv x[i+1]$, however, is the input that the relay would send in the next block $i+1$ had the relay known the state $\dv s[i+1]$, and so is generated at the source using binning against the state $\dv s[i+1]$. The description of vector $\dv x[i+1]$, which is sent to the relay in block $i$, is intended to combine coherently with the source transmission in block $i+1$. In the beginning of block $i$, the relay knows $m_{i-1}$ from the source transmission in previous block $i-1$, and sends $\dv x_2[i]=\hat{\dv x}[m_{i-1}]$.

\begin{remark}\label{remark6}
In the scheme we described briefly in Remark~\ref{remark5}, the relay needs only estimate the code vector $\dv x[i]$ sent by the source in block $i-1$, and transmit the obtained estimate in the next block $i$. For instance, the relay does not need know the state sequence that actually controls the channel. Thus, transmission from the source terminal to the relay terminal can be regarded as that of an analog source which, in block $i$, produces a sequence $\dv x[i+1]$. This source has to be transmitted by the source terminal over a state-dependent channel and reconstructed at the relay terminal. The reconstruction error at the relay terminal influences the rate at which information can be decoded reliably at the destination by acting as an additional noise term.

\end{remark}

\section{The Gaussian RC with Informed Source}\label{secIV}

In this section, we consider a full-duplex state-dependent RC informed source in which the channel states and the noise are additive and Gaussian. In this model, the channel state can model an additive Gaussian interference which is assumed to be known (non-causally) to only the source. The channel outputs $Y_{2,i}$ and $Y_{3,i}$ at time instant $i$ for the relay and the destination, respectively, are related to the channel input $X_{1,i}$ from the source and $X_{2,i}$ from the relay, and the channel state $S_i$, by

\vspace{-0.4cm}

\begin{subequations}
\begin{align}
Y_{2,i}&=X_{1,i}+S_i+Z_{2,i}\\
Y_{3,i}&=X_{1,i}+X_{2,i}+S_i+Z_{3,i}.
\end{align}
\label{ChannelModel__StateDependent__GaussianRC}
\end{subequations}
The channel state $S_i$ is zero mean Gaussian random variable with variance $Q$; and only the source knows the state sequence $S^n$ (non-causally). The noises $Z_{2,i}$ and $Z_{3,i}$ are zero mean Gaussian random variables with variances $N_2$ and $N_3$, respectively; and are mutually independent and independent from the state sequence $S^n$ and the channel inputs $(X^n_1,X^n_2)$.

We shall also consider the following subclass of Gaussian RC with informed source, the parallel Gaussian RC with informed source and orthogonal components where $X_{1,i}=(X_{1R,i},X_{1D,i})$, $Y_{3,i}=(Y^{(1)}_{3,i},Y^{(2)}_{3,i})$ and 
\begin{subequations}
\begin{align}
\label{OutputRelay__GaussianRC__OrthogonalComponents__Parallel}
Y_{2,i} &= X_{1R,i}+S_i+Z_{2,i}\\
\label{OutputDestination__Component1__GaussianRC__OrthogonalComponents__Parallel}
Y^{(1)}_{3,i} &= X_{1D,i}+S_i+Z^{(1)}_{3,i}\\
Y^{(2)}_{3,i} &= X_{2,i}+S_i+Z^{(2)}_{3,i},
\label{OutputDestination__Component2__GaussianRC__OrthogonalComponents__Parallel}
\end{align}
\label{ChannelModel__StateDependent__GaussianRC__OrthogonalComponents__Parallel}
\end{subequations}
where the noises $Z^{(1)}_{3,i}$ and $Z^{(2)}_{3,i}$ are zero mean Gaussian random variables with variances $N_3$, and are mutually independent and independent from the state sequence $S^n$ and the channel inputs $(X^n_1,X^n_2)$. 

A parallel Gaussian RC with informed source and orthogonal components in which the state $S_i$ does not affect transmission from the relay to the destination will be said to be \textit{degenerate}. Its input-output relation is given by \eqref{ChannelModel__StateDependent__GaussianRC__OrthogonalComponents__Parallel} with \eqref{OutputDestination__Component2__GaussianRC__OrthogonalComponents__Parallel} substituted by $Y^{(2)}_{3,i} = X_{2,i}+Z^{(2)}_{3,i}$. 

We consider the following individual power constraints on the average transmitted power at the source and the relay
\begin{equation}
\sum_{i=1}^{n}X_{1,i}^2 \leq nP_1, \qquad \sum_{i=1}^{n}X_{2,i}^2 \leq nP_2.
\label{IndividualPowerConstraintsFullDuplexRegime}
\end{equation}

\subsection{Upper Bound on the Capacity}\label{secIV_subsecB}

\begin{theorem}\label{Theorem__UpperBound__GeneralGaussianChannel}
The capacity of the state-dependent general Gaussian RC with informed source is upper-bounded by\begin{align}
&R^{\text{up}}_{\text{G}} = \max \min \Bigg\{\frac{1}{2}\log\Big(1+P_1(1-\rho^2_{12})(\frac{1}{N_2}+\frac{1}{N_3})\Big),\nonumber\\
&\frac{1}{2}\log\Big(1+\frac{(\sqrt{P_2}+\rho_{12}\sqrt{P_1})^2}{P_1(1-\rho^2_{12}-\varrho^2_{1s})+(\sqrt{\Delta_Q}+\varrho_{1s}\sqrt{P_1})^2+N_3}\Big)\nonumber\\
&+\frac{1}{2}\log\Big(1+\frac{P_1(1-\rho^2_{12}-\varrho^2_{1s})}{N_3}\Big)\Bigg\},
\label{UpperBound__GeneralGaussianChannel}
\end{align}
where $\Delta_Q=QN_2/((\sqrt{Q}+\sqrt{P_1})^2+N_2)$ and the maximization is over parameters $\rho_{12} \in [0,1]$, $\varrho_{1s} \in [-1,0]$ such that
\begin{equation}
\rho^2_{12}+\varrho^2_{1s} \leq 1.
\label{MaximizationRange__UpperBound__GeneralGaussianChannel}
\end{equation}
\end{theorem}
\vspace{0.3cm}

\noindent \textbf{Outline of Proof:} We only sketch the important steps, due to lack of space. The proof of the bound given by the first term of the minimization in \eqref{UpperBound__GeneralGaussianChannel} trivially follows by revealing the state $S^n$ to the relay and the destination. The proof of the bound given by the second term of the minimization in \eqref{UpperBound__GeneralGaussianChannel} is as follows. First, we show that there is an inevitable residual uncertainty at the relay about the state sequence $S^n$ after observing the channel outputs $Y_2^{i-1}=(Y_{2,1},\hdots,Y_{2,i-1})$. Then, considering transmission from the source and relay to the destination, we upper bound the sum rate that can be conveyed to the destination on the multiaccess part of the channel by accounting for the rate penalty that is caused by not knowing the state fully at the relay. In doing so, we assume that the message is revealed to the relay by a genie.

\begin{remark}\label{remark10}
The established upper bound improves upon the cut-set upper bound through the second term of the minimization. The second term of the minimization is strictly tighter than that of the cut-set upper bound because it accounts for not knowing a part $\Delta_S^n$ of the state of power $\Delta_Q$ at the relay.
\end{remark}

\vspace{-0.1cm}
\subsection{Lower Bounds on the Capacity}\label{secIV_subsecC}
\vspace{-0.0cm}


\begin{theorem}\label{Theorem__LowerBound2__GeneralGaussianChannel}
The capacity of the state-dependent Gaussian RC with informed source is lower-bounded by
\begin{align}
R^{\text{lo}}_{\text{G}} = \max \frac{1}{2}\log\Big(1+\frac{(\sqrt{\bar{\gamma}P_1}+\sqrt{P_2-D})^2}{N_3+D+{\gamma}P_1}\Big),
\label{LowerBound2__GeneralGaussianChannel}
\end{align}
where
\begin{align}
D &:= P_2\frac{N_2}{N_2+{\gamma}P_1}
\label{Distortion__LowerBound2__GeneralGaussianChannel}
\end{align}
and the maximization is over $\gamma \in [0,1]$, with $\bar\gamma:=1-\gamma$.\\
\end{theorem}

The proof of Theorem~\ref{Theorem__LowerBound2__GeneralGaussianChannel} follows by using a coding scheme in which the source sends to the relay in block $i$ a quantized version of the input the relay would send in block $i+1$ had the relay known the state of the channel in that block, in the spirit of Theorem~\ref{Theorem__LowerBound2__DiscreteChannel}.

\begin{remark}\label{remar11}
It is insightful to observe that the rate in Theorem~\ref{Theorem__LowerBound2__GeneralGaussianChannel} does not depend on the strength of the state $S$. This makes the described coding scheme appreciable, particularly for the case of arbitrary strong interference for which classical coding schemes have the relay sending at no positive rate when it operates in decode-and-forward mode, because of the unknown interference.
\end{remark}

We now turn to establish a lower bound on the capacity of the state-dependent Gaussian RC using the idea of state description. 


\begin{definition}\label{definition1} Let
\begin{align}
\tilde{Q}_S(t,Q,D) &:=(1-t)^2Q-t(t-2)D\nonumber\\
R(\alpha,P,Q,N) &:= \frac{1}{2}\log\Big(\frac{P(P+Q+N)}{PQ(1-\alpha)^2+N(P+\alpha^2Q)}\Big)\nonumber
\end{align}
for non-negative $t, D, P, Q, N$, and $\alpha \in \mc A(P,Q,N):=\{x \in \mathbb{R}\::\: R(x,P,Q,N) \geq 0\}$.
\end{definition}


\vspace{0.2cm}

\begin{theorem}\label{Theorem__LowerBound1__GeneralGaussianChannel}
The capacity of the state-dependent Gaussian RC with informed source is lower-bounded by
\begin{align}
&R^{\text{lo}}_{\text{G}} = \max\:\min\:\big\{R(\alpha,\beta\bar{\gamma}P_1,\tilde{Q},N_2+{\gamma}P_1), \nonumber\\
&R(\alpha,\beta\bar{\gamma}P_1,\tilde{Q},N_3+{\gamma}P_1)+\frac{1}{2}\log\Big(1+\frac{(\sqrt{\bar\beta\bar{\gamma}P_1}+\sqrt{P_2})^2}{N_3+D+{\gamma}P_1+\beta\bar{\gamma}P_1}\Big)\big\},
\label{LowerBound1__GeneralGaussianChannel}
\end{align}
where
\begin{align}
\label{Distortion__LowerBound1__GeneralGaussianChannel}
D &= Q\frac{N_2}{N_2+{\gamma}P_1}\\
\tilde{Q} &= \tilde{Q}_S(\alpha_2,Q,D)\\
\alpha_2 =& \frac{(\sqrt{\bar\beta\bar{\gamma}P_1}+\sqrt{P_2})^2}{(\sqrt{\bar\beta\bar{\gamma}P_1}+\sqrt{P_2})^2+\beta\bar{\gamma}P_1+(N_3+D+{\gamma}P_1)}
\end{align}
and the maximization is over $\beta \in [0,1]$, $\gamma \in [0,1]$ and  $\alpha \in \mc A(\beta\bar{\gamma}P_1,\tilde{Q},N_2+{\gamma}P_1) \cap \mc A(\beta\bar{\gamma}P_1,\tilde{Q},N_3+{\gamma}P_1)$.
\end{theorem}

\vspace{0.2cm}

\textbf{Outline of Proof of Theorem~\ref{Theorem__LowerBound1__GeneralGaussianChannel}:} An outline of proof of Theorem~\ref{Theorem__LowerBound1__GeneralGaussianChannel} is as follows. The result in Theorem~\ref{Theorem__LowerBound1__DiscreteChannel} for the DM case can be extended to memoryless channels with discrete time and continuous alphabets using standard techniques \cite[Chapter 7]{G68}. For the state-dependent Gaussian relay channel \eqref{ChannelModel__StateDependent__GaussianRC}, we evaluate the rate \eqref{AchievableRate__Theorem__LowerBound1__DiscreteChannel} with the following choice of input distribution. We choose $\hat{S}_D=\O$, $U_D=\O$. Furthermore, we consider the test channel $\hat{S}_R=aS+\tilde{S}_R$, where  $a:=1-D/Q$ and $\tilde{S}_R$ is a Gaussian random variable with zero mean and variance $\sigma^2_{\tilde{S}_R}=D(1-D/Q)$, independent from $S$. The random variable $X_2$ is Gaussian with zero mean and variance $P_2$, independent of $S$ and $\hat{S}_R$. The random variable $X_1$ is composed of two parts, $X_1=X_{1R}+X$, where $X_{1R}$ is Gaussian with zero mean and variance ${\gamma}P_1$, for some $\gamma \in [0,1]$, is independent of $S$, $\hat{S}_R$, $X_2$; and $X=\sqrt{\bar\beta\bar{\gamma}P_1/P_2}X_2+X'$, where $X'$ is Gaussian with zero mean and variance $\beta\bar{\gamma}P_1$, for some $\beta \in [0,1]$, and is independent of $X_{1R}$, $X_2$ and $(S,\hat{S}_R)$. The auxiliary random variables are chosen as
\begin{subequations}
\begin{align}
V &= \Big(\sqrt{\frac{\bar\beta\bar{\gamma}P_1}{P_2}}+1\Big)X_2+\alpha_2\hat{S}_R \\
U &= X' +\alpha (S-\alpha_2\hat{S}_R)\\
U_R &= X_{1R}+\alpha_R(1-\alpha)S
\end{align}
\label{RandomVariables__LowerBound1__GeneralGaussianChannel}
\end{subequations}
where
\begin{subequations}
\begin{align}
\alpha_{2} &= \frac{(\sqrt{\bar\beta\bar{\gamma}P_1}+\sqrt{P_2})^2}{(\sqrt{\bar\beta\bar{\gamma}P_1}+\sqrt{P_2})^2+\beta\bar{\gamma}P_1+(N_3+D+{\gamma}P_1)}\\
\alpha_R &= \frac{{\gamma}P_1}{{\gamma}P_1+N_2}, \:\:\: D := Q\frac{N_2}{N_2+{\gamma}P_1}.
\end{align}
\end{subequations}

\vspace{-0.1cm}
\subsection{Analysis of Some Special Cases}\label{secIV__subsecD}
\vspace{-0.0cm}

\begin{corollary}\label{Corollary__DegenerateParallel__GaussianRC__OrthogonalComponents}
The capacity of the degenerate parallel Gaussian RC with informed source and orthogonal components is given by

\vspace{-0.6cm}

\begin{align}
\mc C_{\text{G-DegParOrth}} = \max_{0 \leq \gamma \leq 1} \min \big\{&\frac{1}{2}\log(1+\frac{{\gamma}P_1}{N_2}),\frac{1}{2}\log(1+\frac{P_2}{N_3})\big\}\nonumber\\
&+\frac{1}{2}\log(1+\frac{{(1-\gamma)}P_1}{N_3}).
\label{Capacity__DegenerateParallel__GaussianRC__OrthogonalComponents}
\end{align}
\end{corollary}

\begin{itemize}
\item[1)] If $N_2 \longrightarrow 0$, the upper bound of Theorem~\ref{Theorem__UpperBound__GeneralGaussianChannel} and the lower bound of Theorem~\ref{Theorem__LowerBound2__GeneralGaussianChannel} tend asymptotically to the same value

\vspace{-0.5cm}

\begin{equation}
\mc C_{\text{G}} =\frac{1}{2}\log\Big(1+\frac{(\sqrt{P_1}+\sqrt{P_2})^2}{N_3}\Big)-o(1)
\label{LowerBound__ExtremeCase__CleanRelay}
\end{equation}
where $o(1) \longrightarrow 0$ as $N_2 \longrightarrow 0$.

\item[2)] If $N_2 \longrightarrow \infty$, the upper bound of Theorem~\ref{Theorem__UpperBound__GeneralGaussianChannel} tends to
\begin{align}
\mc C_{\text{G}} &= \frac{1}{2}\log(1+\frac{P_1}{N_3}),
\label{LowerBound__ExtremeCase__NoisyRelay}
\end{align}
which is achieved by standard DPC at the source.
\item[3)] \textit{Arbitrarily strong channel state:} In the asymptotic case $Q \rightarrow \infty$, the lower bound of Theorem~\ref{Theorem__LowerBound1__GeneralGaussianChannel} tends to
\begin{equation}
R^{\text{lo}}_{\text{G}} =\frac{1}{2}\log\big(1+\frac{P_1}{\max(N_2,N_3)}\big).
\end{equation}
\end{itemize}

\vspace{-0.3cm}
\subsection{Numerical Examples and Discussion}\label{secIV_subsecD}
\vspace{-0.0cm}

Figure~\ref{Fig2__IllustrativeExamples__General__RC} illustrates the upper bound of Theorem~\ref{Theorem__UpperBound__GeneralGaussianChannel}, the lower bound of Theorem~\ref{Theorem__LowerBound2__GeneralGaussianChannel} and the lower bound of Theorem~\ref{Theorem__LowerBound1__GeneralGaussianChannel} for the model \eqref{ChannelModel__StateDependent__GaussianRC} as functions of $\text{SNR}=P_1/N_2$ (in decibels). Also shown for comparison are the cut-set upper bound and the trivial lower bound obtained by considering the channel state as unknown noise and implementing classic DF at the relay. The figure shows that the lower bound  \eqref{LowerBound2__GeneralGaussianChannel} of Theorem~\ref{Theorem__LowerBound2__GeneralGaussianChannel} is asymptotically optimal in $\text{SNR}$. Note that it outperforms the lower bound \eqref{LowerBound1__GeneralGaussianChannel} of Theorem~\ref{Theorem__LowerBound1__GeneralGaussianChannel} for almost all $\text{SNR}$ values. Also, the upper bound \eqref{UpperBound__GeneralGaussianChannel} is strictly better than the cut-set upper bound, as we indicated in the proof of  Theorem~\ref{Theorem__UpperBound__GeneralGaussianChannel}. 

\begin{figure}[!htpb]

        \begin{minipage}[t]{\linewidth}
        \vspace{-0.5cm}
        \begin{center}
        \subfigure[]
        {
	\includegraphics[width=\linewidth,height=0.55\linewidth]{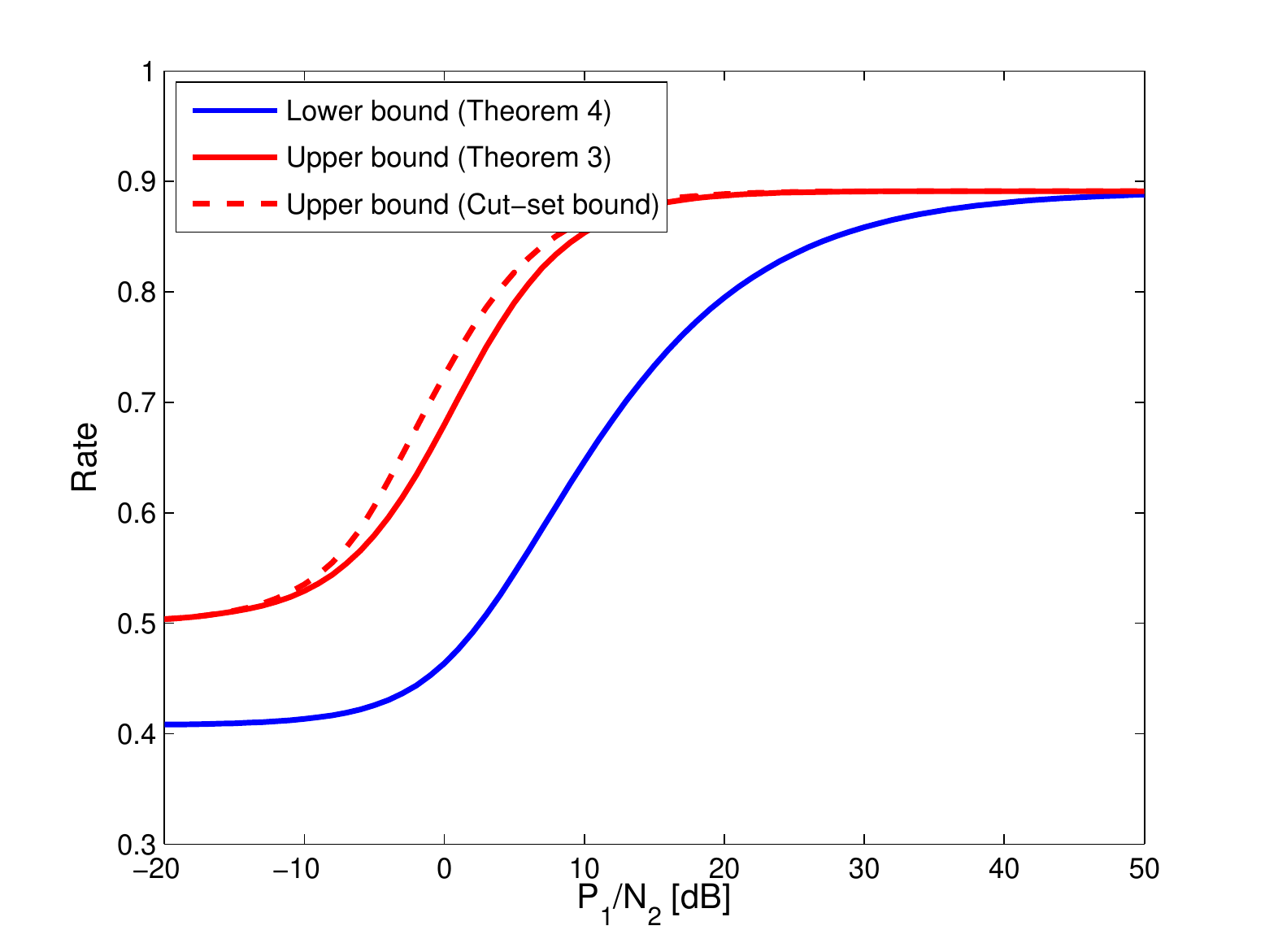}
        \label{Fig2c__IllustrativeExamples__General__RC}
        }
        \subfigure[]
        {
        \vspace{-0.3cm}
        \includegraphics[width=\linewidth,height=0.55\linewidth]{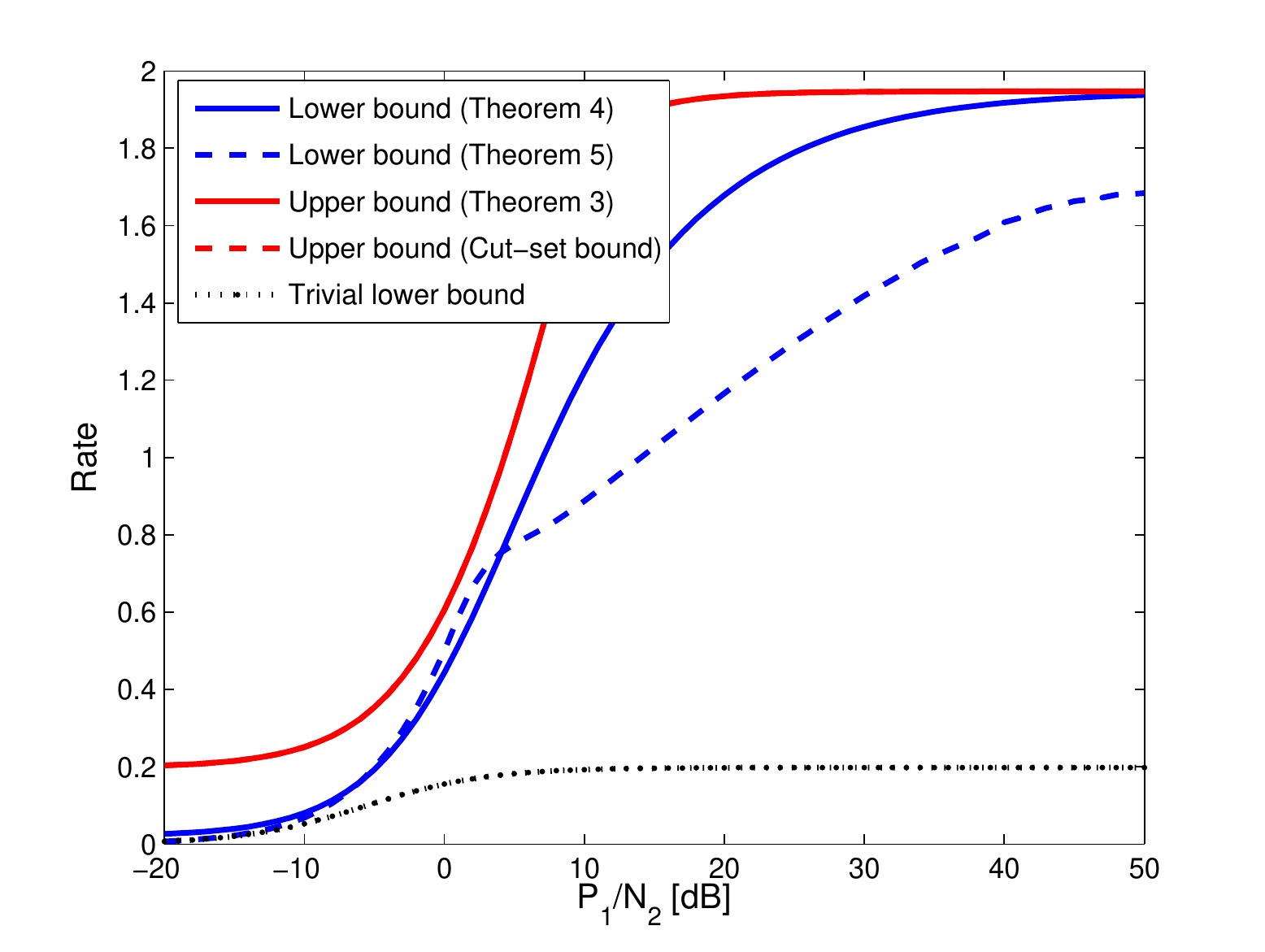}
        \label{Fig2b__IllustrativeExamples__General__RC}
        }
        \vspace{-0.3cm}
        \caption{Lower and upper bounds on the capacity of the state-dependent General Gaussian RC with informed source versus the SNR in the link source-to-relay, for two examples of numerical values (a) $P_1=N_3=10$ dB, $P_2=5$ dB, $Q=30$ dB, and (b) $P_2=40$ dB, $P_1=Q=N_3=10$ dB.}\vspace{-0.3cm}
        \label{Fig2__IllustrativeExamples__General__RC}
        \end{center}
        \vspace{-0.3cm}
        \end{minipage}
\vspace{-0.6cm}
\end{figure}

\vspace{-0.1cm}
\section*{Acknowledgement}
\vspace{-0.0cm}
This work has been supported by the European Commission in the framework of the FP7 Network of Excellence in Wireless Communications. A. Zaidi and L. Vandendorpe also thank the concerted Action SCOOP for funding. The work of S. Shamai has also been supported by the CORNET consortium.

\vspace{-0.2cm}

\bibliographystyle{IEEEtran}
\bibliography{mybibfile}

\end{document}